\documentclass{article}
\usepackage[export]{adjustbox}
\usepackage[utf8]{inputenc}
\usepackage{amsmath}
\usepackage{graphics}
\usepackage[dvipsnames]{xcolor}
\usepackage[english]{babel}
\usepackage[a4paper, total={5in, 7in}]{geometry}
\usepackage{cite}
\usepackage{comment}
\usepackage{dirtytalk}
\graphicspath{ {./fig1.png} }
\usepackage{wrapfig}
\providecommand{\keywords}[1]
{
  \small	
  \textbf{{Keywords:}} #1
}
\title{On the dynamics between gravity and entanglement}
\author{Pradosh Keshav MV \\St. Albert's College\\ Banerjee road, Kochi, Kerala, India\\
 Pin: 682018 
  }
\date{}

\begin{document}

\maketitle
\begin{abstract}
    Recent developments on Bell's experiments demonstrate that entanglement could indeed eliminate the gap between classical and quantum physics. At the same time, it is difficult for a classical theory to include a particular feature like entanglement without compromising the theory's smooth working on a four-dimensional scale geometry. A unified theory should reconsider this difficulty. On the other hand, pregeometry hold the assumption of a non-commutative space where the Requardt-Roy model seems to be a promising one.  From the ordinary five-dimensional approach first initiated by Kaluza-Klein, a toy model is proposed to show the insignificant description of gravity at Planck's scale physics. It is found that the classical nature of quantum correlations are fine tuned within the geometry of space-time at four dimensions. Such a nature can be better understood by studying the pregeometric effects of gravity at five-dimensions. A combined description of gravity and entanglement is found sufficient to explain the fundamental difficulties of discrete space-time manifolds in both the theories.
\end{abstract}
\newpage
\tableofcontents
\newpage
\section{Introduction}

The Suarez-Scarani experiment demonstrates that the non-local correlations cannot be accounted for any past and future (also known as before-before experiment) time-ordered influences on particle pairs\cite{1}. For such a perspective, one has to give up the idea of an event occurring in time before another event that may form the cause of the latter one. Bell experiments generally features an exemplification of the causal notions, as well as physical time which could only make sense in relativistic physics and accordingly one could upholds the orthodoxy Copenhagen interpretation to explain the phenomenon. These experiments are completely contradictory to the relativistic principles and may not hold true in the macroscopic level. In that matter, 'featuring entanglement' as a critical perspective represents the relativistic principles irrespective of their behaviour in the experiment and addresses the non-local feature of it. This is assumed to be done either in four-dimensions or in any $n=(1+(n-1)$ dimensions. One influential five dimensional approach to account for a similar feature (combing electromagnetism with general theory) is from Theodore Kaluza and later initiated by Oskar Klein. In the modelling of a unified theory, they tried to give a plausible explanation to Einsteins field equations inclusive of the electromagnetic phenomenon. But in the long run, there aroused difficulty from the theory to initiate a deviation of electron mass and electric charge ratio from experimental data but the theory does seems to initiate a non-local classical behaviour of a particle\cite{2}. These are achieved by the Scherk-Schwarz gauge symmetry breaking of a d-dimensional field \cite{3}. In turn, the theory could also be addressed by coupling the field constants to an expected value without breaking any local symmetry \cite{4} where I aim to show none of the features were actually wrong from a perspective arsing from the non-local behaviour of particles. \vspace{4mm} \newline
The general framework, therefore, has to be structured from Bell’s theorem since no theory of quantum nature can be weakly non-local. I prefer using the word weakly here since Spekkens\cite{24} formulation of a complete quantum state is more a profound caricature of an existing theoretical presuppositions as argued by Einstein. In that context, I could also comprehend the weakness of a non-local theory based on the Gauge symmetry breaking and categorizing potential sets and functions of Bohmian nature. Moreover, by preserving the principles of relativity in four-dimensions, any non-local feature of an operational theory constitute a comfortable part of General relativity in (4+1) dimensions. Or, in other words, it is never impossible to come with a feature like entanglement whose consequences provided to be a link between classical and quantum theory using extra dimensions. Given that the dimensionality of Hilbert's space itself is infinite, one important aspect to address is the sensitiveness of measurement towards discrete positions of particle pairs that may appear in clock time. Though, it is showed \cite{23} that no macroscopic experiment with massive resolution might address such an issue, regularization of space and time dimension irrespective of any fundamental difference between their norm is one possibility that can be done. But, the problem still arises when the metric fluctuations at the finest coarse grained level make it impossible for the pairs to mutually co-exist at a position less than the Planck's length. Hence, certain assumptions are required to preclude clock time described for a minimal length which is irrelevant without addressing the consequences of finite degrees of freedom in infinite dimensional Hilbert's space. I aim to address this issue by rectifying the regularity of ordinary space-time (minimal uncertainty) using the Requardt cellular network model with minimal norm where two quantum functions $\phi$ and $\psi$ are physically equivalent in very fundamental sense (for a small $\lambda$). Such an attempt using relevant assumptions is evident in the framework of an operational theory that neither deals with relativistic energy as well as quantized electromagnetic field. The Bohmian mechanics, on the other hand, if rectified for relativistic principles in five-dimesnional discrete space-time can be further developed to localize ordinary space-time and produce particle pairs of indistinguishable features. Provided that the five-dimensional formalism may require a framework preceding a critical perspective, I try to substantiate that 'any fundamental assumptions that could possibly link between entanglement and gravity is also a common foundation in classical and quantum mechanics'.

\section{Featuring Entanglement}
It is an intriguing feature of an entangled pairs of electron that at two different directions neither of the pairs can be observed at first. There are about three fundamental quasi-particle assumptions that may contribute to this particular feature of an entangled pair at a distance.\vspace{4mm} \newline 
Assumption 1: The pairs are considered to be non-relativistic based on the scalar hypothesis of Planck and the particle pairs has a probability that integrates to one over all space and that the particle is not created nor destroyed\footnote{The probability density over a small density of space i.e, $d^3 x$ is one if $\int (\mid \psi \mid^2 )d^3 x = 1$ which is called the normalization condition.}. With a setup, whenever the measurement is made on one pair at ‘t = 0’, for e.g. the spin measurement, there also correlates a set of information such that the function at rest position $f(R)$ of particle is commuted irrespective of their spatial temporal property. Hoshang Heydari showed that the space of a pure entangled state is a non-commutative space \cite{5} but there are no sources of the field that contribute to the feature here (the equivalent of charges and currents in electromagnetism). If we assume a scalar five-dimensional quantity $\zeta(R)$ where R represents a flat space-time, then the source term of the quantity also be a scalar that bundles with the stress-energy term $T$ of the source density. A non-relativistic particle interaction can be addressed in a five dimensional space-time such that in a four dimensional space-time, the principles of relativity are never violated. Moreover, with a spontaneous precision irrespective of their distances in space, from Assumption 1, an indication of an absolute “now” is essential to deduce any local effect of an entangled pairs of electrons. Here, $\tau$ is not to be confused with the physical time, which is expected to be an emergent coarse grained quantity.\vspace{4mm} \newline
Assumption 2:  The position of the two pairs independent of physical time is reduced from a field symmetric component\footnote{With the assumed symmetry of pseudo-Riemannian R, the principle of least action gives the Noether's theorem. The classical intuition suggests that non-invasive measurements should be time symmetric where  the disturbance becomes negligible as the measurement strength goes to zero\cite{6}. This is not possible in quantum level since non-invasive measurements cannot be time-symmetric in quantum weak measurements.} of R based on assumption 1. There requires an arbitrary extra dimension\footnote{One claim with experimental evidence\cite{7} suggests that there is a time varying fine structure constant $\alpha = \frac{e^2}{(4 \pi \hbar C )}$ from where I am expecting a postulate that $\hbar C$ changes in time. If that is the case, the consequences will be a breaking of Lorentz invaraince or conservation of energy where both has to be excluded for a choice to keep assumption 2 of exclusive authenticity. If there is a changing $\hbar$ and constant C, one way I could approach $\alpha$ is by structuring an arbitrary dimension of time where different dimensionless entities interacts in a five-dimensional field uni-directionally and bi-directionally somewhere before and now in time. This arbitrary dimension of time is then inclusive of all superluminal structures and particle's energy-momentum conservation laws.} of time $\tau$ whenever the particle interact with an external body at a particular or featured time such that the particle from its rest position is then inclusive of all functions of $\zeta(R)$ and its spin eigen function represents the field symmetry between the two positions. \vspace{4mm}\newline
If $H_{\alpha \beta}$ is a four vector 'Real space' that has three dimensional Real linear space $E_{\alpha \beta}^{3}$ , then the four components $(ct, x_1, x_2, x_3)$ can be expressed in a matrix form of QP.
\begin{equation}
    QP =
    \left(\begin{matrix}
ct & -\alpha x_1 & -\beta x_2 & -\alpha \beta x_3 \\
x_1 & ct & -\beta x_3 & \beta x_2 \\
x_2 & \alpha x_3 & ct & -\alpha x_1 \\
x_3 & -x_3 & x_1 & ct
\end{matrix}\right)
\left(\begin{matrix}
ct' \\
x_{1}'\\
x_{2}'\\
x_{3}'
\end{matrix}\right)
\end{equation} Where the complex conjugate of Q and P yields,
\begin{center}
  \begin{math} 
  QP \ =\ QP +\ \zeta (Q P^* + Q^* P) 
  \end{math}
\end{center}
\begin{center}
    \begin{math}
     PQ \ = \ PQ +\ \zeta (PQ^* + P^* Q) 
    \end{math}
\end{center}
Which essentially shows that the Real space is a \textit{dual quoternionic} space. As a part of eq(1), the four vector language can be used to formulate the quoternionic space $H_{\alpha \beta}$ where the Real space interval between Q and P is a non-commutative space interval. Thus, for the inner products,
\begin{equation}
   Q \ = \ A_0 + A_{1} \hat i +  A_{2} \hat j +  A_{3} \hat k 
\end{equation} and
\begin{equation}
 P \ = \ B_{3} \hat k + B_{2} \hat j + B_{1} \hat i + B_{0}
\end{equation}
If $R \ = Q P$ , then $QP \neq PQ$ which gives,
\begin{center}
    \begin{math}
   R = A_{0}B_{0} - \alpha A_{1}B_{1} - \beta A_{2}B_{2} - \alpha \beta A_{3}B_{3} + (A_{0}B_{1} + A_{1}B_{0} - \beta A_{2}B_{3} + 
 \beta A_{3}B_{2} )\ \hat i + (A_{0}B_{1} + A_{1}B_{0} - \beta A_{2}B_{3} + \beta A_{3}B_{2} ) \ \hat j + (A_{0}B_{3} + A_{1}B_{2} - A_{2}B_{1} + A_{3}B_{0} ) \ \hat k 
    \end{math}
\end{center} We now get a metric component of $R \ = QP$ for all $\alpha \beta \in H$, a set of all generalized quoternions.
\begin{equation}
    QP =
    \left(\begin{matrix}
A_{0} & -\alpha A_1 & -\beta A_2 & -\alpha \beta A_3 \\
A_1 & A_{0} & -\beta A_3 & \beta A_2 \\
A_2 & \alpha A_3 & A_{0} & -\alpha A_1 \\
A_3 & -A_2 & A_1 & A_{0}
\end{matrix}\right)
\left(\begin{matrix}
B_{0} \\
B_{1}\\
B_{2}\\
B_{3}
\end{matrix}\right)
\end{equation}
And the dual space has the form 
\begin{center}
    \begin{math}
     H_{\alpha \beta} \ = \ R \ \oplus \ E_{\alpha \beta}^3 
     \end{math}
     \end{center}
     \begin{center}
     \begin{math}
    \Bar{H}_{\alpha \beta} \ = \ \Bar{E}_{\alpha \beta}^3 \ \oplus \ R' 
    \end{math}
\end{center} 
Where R and R' has a set of positive real numbers.\vspace{3mm} \newline
The signature of a five-dimensional metric with three space dimensions and two-time dimensions can be referred from equation\( ~as~~ \left(+\ +\ +\ +\ - \right)  \) \textbf{.}  With the simultaneity of investigating multiple dimensions of time, we can define a mass-less scalar field  \(  \phi = \left(  x_{1},x_{2},x_{3},ix_4, \tau \right)  \)  described by a wave equation. The generalization of wave equation with $n$ dimensions and mass $m$ is a  Klein-Gordon equation.
\begin{equation}
 [\partial_{i} \partial^{i}- m^2]\phi = 0 
\end{equation}
It is also possible to consider similar model from a free complex scalar field $\phi$ satisfying eq(5). In \cite{8} V. Balakumar and E. Winstanley  have made a generalization of eq(5) to consider a charged complex scalar field on a general d-dimensional curved space-time background , in that case the partial derivatives
$\partial_{i}$ are replaced by space-time covariant derivatives $\delta_{i}$. Then the equation take reals values and describes the propagation of components of $\zeta$  function. The generalization of the wave equation with three space and two-time dimensions is the free complex scalar field $\phi$ satisfying eq(5).
\begin{equation} 
 [\delta_{i} \delta^{i}- m^2-\zeta (R)]\phi  = 0 
\end{equation}
It is useful to write the square distance between the two neighboring points $\delta(x)$ and $\delta(x+\delta x)$, in the line element, in reference frame $s(x)$ as,
\begin{center}
\begin{math}
    ds^2 = d\delta^i \ \zeta _{ij}(x) \ d\delta^j 
   \end{math}
\end{center} 
Assumption 3: There is a possible set of commuting and non-commuting observables based on Bohm's hypothesis\footnote{As one Bohmian is considered of, the guiding wave feature of quantum interpretation seems to be in conflict with the causal feature of quantum potential. Though Bohm was convinced by this feature at the beginning, his later works with Basil Hiley gradually negated this feature where uncertainty relations are given an epistemic understanding rather than an ontic one \cite{9}. Moreover, Bell theorem proved that every system in agreement with the observed probabilities of quantum theory should follow the predictive rules of quantum theory which also concludes that every system (including a Bohmian system) following these rules should also be non-local \cite{10},[7-8].}of causal particle feature combined to form a set of commuting operator whose eigenvalues define the source strength of an entangled system. The source density can be built up from point sources so it is useful to understand the time independent field generated by a point source as we do for electromagnetism \begin{math}
 \zeta( x,t)\ =  \zeta( R) =  G \ \delta ^{i}(x) 
 \end{math}; this is a source of strength G at the origin. $\zeta(R)$ does not change with time so we expect a static scalar field. The resultant field is the sum of wave energy and particle momentum represented in a simple tensor bundle as
\begin{equation} 
\zeta~T_{i}^{jk}  =  ~\zeta_{i} \nabla _{ij} \int _{}^{}2 \pi  \nu  \frac{8 \pi G}{C^{4}}~ \delta ^{i} \delta ^{j}  +   \int _{}^{} \hbar r\omega\sum _{i=1}^{n}\frac{m_{i}}{2 \pi }
\end{equation} Where I use the stress-energy term $T$
replaced by a space-time derivative $\delta$ to generate a reduced  Einstein's constant $K$ in addition to induced Planck's mass $m$ for a five-dimensional curvature scalar R and mass-less scalar field $\phi$ where 
\begin{equation}
\nabla = \frac{\partial}{\partial_x}-A
\end{equation} The resultant complex quantity $A=iKr$ of any finely tuned particle is then expected to satisfy the  Einstein's field equations for $R$ in terms of Planck's mass $m$ and effective Planck's wavelength of $r$ with a minimal uncertainty\footnote{There is supposed to be a lower limit on the product of the uncertainty of position and momentum such that the lower bound applies only after the measurement. Indeed, one could also measure the momentum of the particle with arbitrary precision prior to measuring its position with arbitrary precision \cite{11},[84-87].} arising from eq(6) as expected form eq(7) that can be expressed in terms of $\delta (x)$ and momentum $\delta (p)$ in regard with the source density and plank's constant as
\begin{equation}
   G \frac{h  r}{4 \pi } \leq  [\delta(x), \delta (p)]
\end{equation}
But here, eq(9) should be written in polar form for a Bohmian system to satisfy the real values of R(x,\ t) and S(x,\ t) by rewriting the Schrodinger's wave equation 
\begin{equation}
 {\psi = R\ exp^({\frac{i S}{h}})}
\end{equation}
An identical non-relativistic wave equation also looks similar from eq(7) as 
\begin{equation}
 {\psi = \int \zeta_{i}\ exp^(\frac{\hbar r^{3}}{2\pi}}) \ d(\nabla_\zeta)
\end{equation}
From eq(11) and eq(10) a reduced Planck's constant can be compared and formed by substituting the polar form of r and S to eq(9) as an indication of minimal uncertainty arriving from Klein-Gordan equation, 
\begin{equation}
 \hbar r  = \sqrt{\frac{iS}{r}} 
 \Rightarrow \ G \frac{1}{2} \sqrt{\frac{iS}{r}} \leq  [\delta(x), \delta (p)] 
\end{equation}
Based on these three assumptions, there are a set of five-dimensional quantities we can come up with in order to understand the non-locality of entangled pairs of particles. The prescribed four dimensions quantities from eq(4) then should satisfy the linear vector space in its Real form as,
\begin{equation}
     R^5 \ = \ C^4 \times S_1 
\end{equation} Where $S_1$ yields a cube. This is similar to the five dimensional approach of Kaluza-Klein model which is detailed in section 3.1.

\section{Action at a minimal distance}
In all preogemetric structures, it was Requardt cellular networks that went beyond the manifold assumption by means of a graph-theoretical approach. A law for the evolution in what he called the “clock-time” of such a graph is introduced on the discrete grounds that it provides, by trial and uncertainties, some desired consequences \cite{12}. Requardt and Roy  assumed that this graph should have evolved from a chaotic initial phase in the distant past, characterized by the complete absence of stable patterns where it eventually reaches a stable pattern in the context of ordinary space-time. In such a way, the underlying substratum of Planck's scale physics or, more specifically the space-time vacuum (voids) can be viewed as cellular network\cite{13}. This discrete structure consists of elementary nodes $\zeta_{i}$ which interacts with each other via $\zeta_{jk}$ bonds playing a difficult role to reduce generators through an elementary interaction. The possible internal structures of the nodes, as described by Sisir Roy is a discrete internal tensor space carried by the nodes itself. The internal state of the nodes/bonds are denoted by generators $
\zeta_{j} \zeta_{k}$ and  $\zeta_{k} \zeta_{j}$ respectively. My aim is thus to extrapolate a complex interactive field out of Hamilton-Jacobi equation.\newline
A massive scalar field falls off exponentially and larger the mass, faster the fall off\footnote{At Planck's scale, space-time has a granular structure. By investigating in to a massive scalar field in "n" scales that are generalizations of time scales into higher dimension, F. S. Dunder found that field equations generated from the model is complex and only be solved through computer simulations\cite{14}.}. So from eq(7), the generator $\zeta_{i}$ demands an extrapolation in terms of Plank’s time-space quanta as, 
\begin{equation}
     \zeta_{i} \Rightarrow \zeta \zeta_{j}+\zeta \zeta_{k}+\zeta_{j}\zeta_{k}+\zeta_{k}\zeta_{j} 
 \end{equation}
Any arbitrary rational function $\zeta_i$ can also be represented as a tensor space derivative $\nabla (x,\ x')$ between two masses separated by $r$. Angular wavelength that previously substantiated the tensor space of $\nabla$ over a cubic geometry is taken as $\frac{\hbar r^{3}}{2 \pi }$ from eq(11) where the power exponential function of Planck’s effective $r$ bonds with $\zeta_{j} \zeta_{k}$ and transforms into space exponential function in Hilbert’s space $H$ by Savitch theorem\footnote{In computational complexity theory, Savitch theorem confronts the relationship between deterministic and non-deterministic space\cite{15},[135-138]. For all r,n,1$\leq$ r $<$ n, there is a set "a" of strings which has non-deterministic time complexity $n^n$ but not non-deterministic time complexity $n^r$ \cite{16},[187-192]. }. In $n$ dimensions, for any function $f$ Savitch theorem confronts that,
 \begin{center}
 \begin{math}
 f  \in \psi (log(n)) 
 \end{math}
\end{center}
\begin{center}
\begin{math}
 \Rightarrow NSPACE (f(n)) \subseteq DSPACE ((f(n))^2) 
 \end{math}
\end{center}
 If $\zeta_i$ is context free language, then
\begin{center}
\begin{math}
\Rightarrow exp^(\frac{\hbar r^{3}}{2\pi}) \int \zeta _{i}\ d(\nabla _{\zeta}) \ \in \ H  
\end{math}
\end{center}
By expanding the term we get,
\begin{equation}
 exp^(\frac{\hbar r^{3}}{2 \pi }) \int ( \zeta \zeta_{j}+\zeta \zeta_{k}+\zeta_{j}\zeta_{k}+\zeta_{k}\zeta_{j})\ d(\ \hat{i}\frac{\partial}{\partial_{\zeta}}+\hat{j}\frac{\partial}{\partial_{\zeta}}+\hat{k}\frac{\partial}{\partial_{\zeta}} - iKr)
\end{equation}
This is a five dimensional field satisfying the signature matrix (+\ +\ +\ +\ -) from assumption 2.\vspace{5mm} \newline
If $Q_{i} = Q_{i} (q,p)$ and $P_{i}= P_{i}(q,p) $ is without explicit dependence of time, the transformation is restricted to canonical where $f$ depends on mix of old and new phase space variables; it is called a generating function of canonical transformation. 
i.e. if $f = f_{1}(q,Q,t)$ where the old coordinates $q_{i}$ and new coordinates $Q_{i}$ are independent, then for Hamilton $H_{e}$ we can be find the value of K.
\begin{equation}
p_{i}\dot{q_{i}}- H_{e}=P_{i}\dot{Q_{i}}+\frac{\partial f_{1}}{\partial_{\zeta}}+\frac{\partial f_{1}}{\partial{q_{i}}}\dot {q_{i}}+\frac{\partial f_{1}}{\partial{Q_{i}}}\dot {Q_{i}}-iK\frac{\partial r}{\partial t}
\end{equation}
which gives,
\begin{equation}
    H_{e}=iK\frac{\partial r}{\partial t} - \frac{\partial f_{1}}{\partial_{\zeta}}
\end{equation}
where the Hamilton-Jacobi equation is expected to give the  semi-classical limit of quantum mechanical Bohmian system from eq(10) with a standard $p^{2}/m$ kinetic term in $H_{e}$. Here a change in variable is made without a loss in generality and S(q,t) is complex.
\begin{equation}
   -\frac{\partial{S}}{\partial{t}} = \frac{i\hbar}{2m}\frac{\partial^2{S}}{\partial{q}^2}+\frac{1}{2m}(\frac{\partial S}{\partial q})^2  +V(q) 
\end{equation}
If we take $\hbar \rightarrow
\frac{1}{r}\sqrt{\frac{iS}{r}} $ the imaginary term will go away from eq(13).
\begin{equation}
      -\frac{\partial{S}}{\partial{t}} = \frac{1}{2mr}\sqrt{\frac{S}{r}}\frac{\partial^2{S}}{\partial{q}^2}+\frac{1}{2m}(\frac{\partial S}{\partial q})^2 + V(q)
\end{equation}
then from eq(17)
\begin{equation}
    0=\frac{\partial{S}}{\partial{t}} + H_{e}(\delta (x),\frac{\partial S}{\partial \delta})
\end{equation}
This is the modified Hamilton-Jacobi equation of a Bohemian system. The classical action is put together by an indefinite integral over $\zeta$ of the Lagrangian $L$.
\begin{center}
\begin{math}
 \dot{S} = p_{i}\dot{\delta_{i}} - H_{e} = L  \Rightarrow S=\int L \ d \zeta
\end{math}
\end{center}
This is an incredible classical action at a minimal distance where we can compare and contrast the non-relativistic wave equation from eq(11) with that of the classical Lagrangian and find the value of $r$ and reduced $K$. \begin{equation}
  \int L \ d \zeta \ \simeq \int \zeta_{i}\ exp^(\frac{\hbar r^{3}}{2\pi})\ d(\nabla_\zeta)
  \end{equation}
\begin{center}
\begin{math}
     \Rightarrow p_{i}\dot{\delta_{i}} - H_{e} \ \approx \ \zeta_{i}\ exp^(\frac{\hbar r^{3}}{2\pi}) 
     \end{math}
\end{center}
\begin{equation}
 \Rightarrow iK\frac{\partial r}{\partial t} - \frac{\partial f_{1}}{\partial_{\zeta}} \ \approx \ p_{i}\dot{\delta_{i}} - \zeta_{i}\ exp^(\frac{\hbar r^{3}}{2\pi})
\end{equation}
From eq(22) we can approximate the values,
\begin{center}
\begin{math}
  \frac{\partial f_{1}}{\partial_{\zeta}} =   \zeta_{i}\ exp^(\frac{\hbar r^{3}}{2\pi})\ and \ iK\frac{\partial r}{\partial t} = p_{i}\dot{\delta_{i}}
  \end{math}
\end{center}
Which gives an arbitrary complex equation of r as
\begin{equation}
    r = \sqrt{2 \pi \sqrt{\frac{r}{iS}}\ log_{\zeta_i}(\frac{\partial f_{1}}{\partial_{\zeta}})} 
\end{equation}
 Solving for r and K in eq(15) we get the minimal action $S_a$ as,
 \begin{equation}
  S_a= \int_{a} \zeta \ \frac{m \hbar}{2 \pi C^2} + [\epsilon_r,  \epsilon_n] \ F [\sigma_r ,\sigma_n] 
 \end{equation}
where $[\epsilon_r , \epsilon] _n$ is a set of commuting observables and $[\sigma_r, \sigma_n]$ is a set of non-commuting observables. The minimal action at $S_{a}$ shows that the tensor space is replaced by a class of strings on this space. This is similar to the philosophy of non-commutative geometry as pointed out by Sisir Roy. The natural distance function is given by $d (\zeta_{j},\zeta_{k})$ where the bond $b_{ik}$ is assumed to connect the nodes $\zeta_{j}$ and $\zeta_{k}$ and  bond $b_{ij}$ is assumed to connect the nodes $\zeta_{k}$ and $\zeta_{j}$ i.e., the minimal action connecting $\zeta_{i}$ is also the effective length r of a path.

 \subsection{Effective functors and Beyond Planck's Field}
 A tangential approach using a five dimensional metric is comparatively an older one where the popular is from Theodore Kaluza. He had proposed a five dimensional theory of relativity, so that one could geometrically unify gravity and electromagnetism through an effective action of a five dimensional Ricci scalar \cite{17}. Further, he imposed the cylinder condition, i.e. the components of the metric $g_{ij}$ should not depend on the space-like fifth dimension which is an important assumption for his theory. A similar approach is made in this paper by reconstructing the Kaluza-Klein action in $ 5= (1+(5-1))$ which will describe a massless scalar field and pure gravity.
 \begin{equation}
     S_{kk}= S_a+S_e= \alpha_1 \int \frac{1}{r} \sqrt{\frac{r}{iS}} + [\epsilon_r, \epsilon_n] \ F [\sigma_r ,\sigma_n] - \alpha_2 \int d^5xER^5
 \end{equation}
 Where $S_a$ is the minimal action, $S_e$ is the Einsteins action, $\alpha_1$ is the bond coupling constant
$\zeta \frac{m_p}{2\pi C^2} $ , $\alpha_2$ is the gravitational coupling constant
 $ G\frac{ m_e^2}{\hbar C} $ and $ E=\sqrt{-\Phi}$. The scalar R from eq(6) gives a straight forward calculation of Ricci scalar as, \begin{equation}
     R^5 \ = \ R - \frac{1}{4} \phi F_{ij}F^{ij} - \frac{2}{\sqrt{-\Phi}} \partial_i \partial^j \sqrt{-\phi}
 \end{equation} 
 Here the Ricci scalar R can further be defined using a functor, 
 \begin{equation}
     R=F^{r \ [^a F^{nb}]} (\partial_n \sigma_r - \sigma_n)
 \end{equation}
where [...] (square bracket) from eq(27) implies anti-symmetrization and allows morphisms of $\zeta$ functions for all \begin{center}
 \begin{math}
    \sigma_r \sigma_n < f(\zeta)_a, f(\zeta)_b
 \end{math}
 \end{center}
 which gives,
 \begin{center}
 \begin{math}
     F^{r \ [^a F^{nb}]} = F^{na} F^{rb} - F^{nb} F^{ra} 
 \end{math}
 \end{center}
In category theory, a functor F from a category C to category C' is defined as : for each object X in C , F assigns an object F X in C', such that F assigns a morphism 
\begin{center}
\begin{math}
  Ff: FX \rightarrow FY \in C'
\end{math}
\end{center}
 Then F must satisfy the identity morphism $id_x$ as,
\begin{equation}
     F id_x = id_{FX} 
\end{equation}
Now let the Hilbert space $H$ be a topological space or a Euclidean $R^n$ space. Then $\hat H$ is the category of contravariant functors from the category C associated with a topological space H to a product category $\prod_{e\in \Gamma} C_{e}$ where $\Gamma$ is an index set and e is the basis vector. The category C is said to be the generalized time space or generalized time category when the real value of $\zeta$ is embedded in $H$.
\begin{equation}
    \hat H = \prod_{e \in \Gamma}^n C_{e}^{H^{opp}} 
\end{equation} where $H^{opp}$ is called the dual category of $H$ such that $(H)^{opp}$ = $H$. The concept of dual category enables to dualize each notion with respect to a category H into another notion with respect to the category $H^{opp}$.
Thus the five dimensional space-time has the topology $C^4 \times S_1$, where $C^4$ is 4-dimensional Minkowski space-time and $S_1$ is a cube.
\begin{figure}[ht]
    \centering
     \includegraphics[scale=.55]{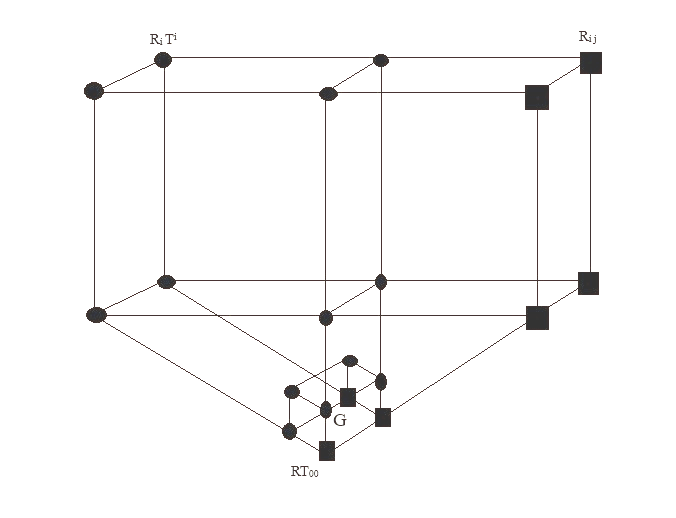}
       \caption{Between two category of a topological space has bonds illustrated as multiple cubes with source density at the origin.}
     \label{fig:my_label}
\end{figure}
So far in general, from the action $S_{kk}$ where \begin{math}
\alpha_1 = V \hat \alpha_1
\end{math}
and \begin{math}
\alpha_2 = V \hat \alpha_2
\end{math} (hat signifies a five-dimensional quantity) we assume that $\frac{\alpha_1}{\alpha_2}$ = 1. Solving for the term $\zeta$ we get \begin{equation}
     \zeta = G \ 2\pi \frac{m_e^2}{m_p} 
\end{equation}
This equations tells the direct relation between $\zeta$\ and source density G where we can signify how the nodes are connected to G in Planck's scale, where $\hbar$ = C = 1. If we assume that $\mu = \frac{m_e^2}{m_p} = 1$ then we get a direct relation $\zeta = 2 \pi G$ and the minimal action at eq(24) turns into,
\begin{equation}
   \frac{2 m_p}{C^2} \int \frac{G}{2}\sqrt{\frac{iS}{r}} + \epsilon_r \epsilon_n \ F [\sigma_r ,\sigma_n] 
\end{equation} From eq(12), we can show that for all $x$, $p$ $\in$ $H$ and r, n $\in$ $H^{opp}$ where n= (1,2,..) 
\begin{equation}
     [\delta (x), \delta (p)] \geq Fid_{\epsilon} [\sigma_r, \sigma_n] \geq \frac{\hbar}{2}
\end{equation} This is a generalized quantity satisfying eq(29) and a simplified model of a five-dimensional Kaluza-Klein action $S_{kk}$. The model also represents a simplified Planck's mass interaction at the nodes $\zeta_{i}$ bonded to categories of $\hat H$ which is a complex function of contravariant functors. A similar category that trigger the coupling constants $\alpha_1$(the minimal action) and $\alpha_2$(Einsteins action) can be called as Planck's action at effective wavelength $r$ from eq(23). The five-dimensional Bohemian system from eq(25) obtained from the Planck's action can be called for simplicity beyond Planck's action, or Beyond Planck's Field (BPF) that is illustrated as a cube with notion G in the figure 1. The BPF invariance illustrated in figure 1. for any transformation is appreciable under the category of contravariant field tensors $T^i$ and covariant tensor $R_{i,j}$  such that no two indices i,j or j,k or i,k transform under a radical symmetry. Gauge symmetry coming out of this distinguished preogeometric space-time is called BPF scalar R (a pseudo-Riemannian scalar) which can further demonstrated through identity morphisms $id_\epsilon$ of Planck scale Functor FX. Moreover the radical field strength coming from the gauge symmetry of Planck's effective r could be determined by the mass invariant transformation  given in eq(26) and further yields the possibility of applying Einstein’s field equations for a five-dimensional flat space-time.   
\section{Invariant mass transformation and Scale symmetry} 
\subsection{what concerns the scalar field curvature}
It is a made fact that from General relativity, any mass could transform and bend the fabric of space-time. Since our understanding about larger masses are well defined with Einstein's four-dimensional field equation, I also finds a necessity to include the BPF scalar in to a well defined Einstein's field equations. As follows, adding an extra dimension of space to a four-dimensional metric is not an elegant way of representing the plank scale functor dynamics. So, I adopt an extra dimension of time to include any feature including entanglement for a scale symmetric transformation (SST). Recent works from Rochester researchers showed that Bell's inequality is not an actual boundary between the classical and quantum world. They confronts that Entanglement could actually fill this gap as a bridge between the real and quantum world\cite{18}. According to Eberly, every system irrespective of quantum or classical should be ready to tackle this feature. Classical entanglement which was introduced in the late 1980s couldn't explore the phenomenon much since the effect was all local. In this section, I argue that locality as we have understood from the Einsteins equations could impose an opposite effect on this thinking. Locality has only explored in classical mechanics with regulations kept to impose reality as a four-dimensional metric. Perhaps, in five-dimensional, I suppose there can be geometric effects turning into pregeometric effects with radical scale symmetries achieved by the presupposition of ordinary quantum theory. How could a non-relativistic theory explore this feature so to check the readiness of the scale symmetries from massive scales to effective Planck's scales? One way I could signify this feature is by checking the compatibility of Einstein's Field equation to accompany effective masses (transformation of tensor variables apparently with a indefinite Planck's mass) interconnected by nodes in a flat space-time. \vspace{3mm} \newline
From assumption 3, we have considered a Bohmian deterministic system that goes well with a scalar action in five-dimensions. To achieve an invariant mass transform, the contribution of a constant is necessary such that the modified scalar function could replace the original Einstein's field equations without perturbations that may contribute to the expansion of the metric. Assuming that the curvature is flat and R represents the scalar, the space exponential function is considered to be a static with $\Lambda$ constant. The four-dimensional equation appears to be,
\begin{equation}
  R_{ij} - \frac{1}{2} g_{ij} + \Lambda g_{ij} =- f(\zeta)_a 
\end{equation}
The $'-'$ sign indicates that for all Planck scale functors, the geometry cant be determined by the tensor bundle beyond a conserved scale symmetry. Thus the following symmetry is allowed,
\begin{equation}
     \zeta R_{k} + \frac{i^2}{2} f(\zeta_k) - \zeta_{jk} \int f(\zeta_i) \frac{\hbar r^3}{2\pi m} = K S_a 
\end{equation}
Including the corrected term for $\hbar$ and minimal action $S_a$ we get,
\begin{equation}
      \zeta R_{k} + \frac{i^2}{2} f(\zeta_k) - \zeta_{jk} \int f(\zeta_i) \frac{r^2}{2 \pi m} \sqrt{\frac{iS}{r}} = \frac{8 \pi G}{C^4} \int_{a} \zeta \ \frac{m \hbar}{2 \pi C^2} + [\epsilon_r , \epsilon_n] \ F [\sigma_r ,\sigma_n] 
\end{equation}
where the RHS of the eq(35) represents the zero point energy and BPF tensor generators which is represented by functions of n dimensional manifolds and a cube of effective Planck's length r. The LHS represents the original Einstein's field equations with a modification of Bohemian deterministic $\zeta$ functors. One can actually derive eq(20) (quantum potential $Q(R)$ and classical potential V(q)) by limiting the conditions of cube r = $l_p$ and $m = m_p$ where $l_p$ is the Planck's length and $m_p$ is the Planck's mass.
\subsection{Gauge symmetry and invariance of vector products}
Let M be a (n+1)-dimensional Minkowskian space-time with metric $g_{\mu \nu}$, the metric induced on R is,
\begin{equation}
  \hat g_{\mu \nu} =
    \begin{Bmatrix}

      g_{\mu \nu} & g_{\mu 5}\\
      g_{\nu 5} & g_{5 5}
    \end{Bmatrix}
\end{equation}
 The 5th dimension is postulated to be compactified, modelled into a small cube, which provides us the explanation for the inconceivability of the non-abelien group 
 \begin{math}
 \Sigma \ [r_{1, \dots N},\ n_{1, \dots 5}] 
 \end{math} belongs to the category $H^{opp}$.  The general form of M is $C^4 \times S_1$,
 \begin{equation}
 \hat g_{\mu \nu} = g_{\mu \nu} + \partial_\mu f \partial_\nu f 
 \end{equation}
The first component of metric tensor can be obtained from basis vectors $e_\mu$ and $e_\nu$ as
\begin{equation}
   g_{\mu \nu} =  e_\mu * e_\nu = g_{\mu \nu}^4 + \frac{g_{\mu 5}g_{\nu 5} }{g_{5 5}} 
\end{equation} where '*' represents the invariance of vector products. By investigating into the commutator $[XX]$ of Pauli's spin matrix and their Kronecker product * we get,
\begin{equation}
     \zeta * R_k \rightarrow X_{\alpha \beta} + XX 
\end{equation} The possible values of $\alpha$ and $\beta$ are
\begin{gather*}
      X_{0 0} + XX \\
    X_{1 0} + XX \\
   X_{0 1} + XX \\
   X_{1 1} + XX 
\end{gather*}
 and $\sigma_r$ $\sigma_n$ takes the values,
 \begin{center}
 \begin{math}
 \sigma_r^* =
 \left(\begin{matrix}
0 & 0 \\
1 & 0 \\
0 & 1 \\
1 & 1 
\end{matrix}\right)
\end{math} \ \ and \ \ \begin{math}
\sigma_n^* =
\left(\begin{matrix}
0 &  0 \\
-1 & 0 \\
0 &  1 \\
1 & -1
\end{matrix}\right)
\end{math}
\end{center}
where,\begin{equation}
    i^* = \sigma_r^* + \sigma_n^* \rightarrow
        \left(\begin{matrix}
0 & 0 \\
0 & 0 \\
0 & 2 \\
2 & 0
\end{matrix}\right)
\end{equation} Four dimensional gauge symmetry with * represents the complex Pauli $j^*$ and $k^*$ matrices as,
\begin{equation}
    j^* \rightarrow
    \left(\begin{matrix}
0 & 0 & 0 & 0 \\
1 & 0 & 1 & 0 \\
0 & 1 & 0 & 1 \\
1 & 1 & 1 & 1
\end{matrix}\right) \times
\left(\begin{matrix}
0 & 0 & 0 & 0 \\
0 & i & 0 & 1 \\
1 & 0 & -i & 0 \\
1 & 1 & 1 & -i
\end{matrix}\right) =
\left(\begin{matrix}
0 & 0 & 0 & 0 \\
1 & 0 & -i & 0 \\
1 & 1+i & 1 & 1-i \\
2 & 1+i & 1-i & 1-i
\end{matrix}\right)
\end{equation}

\begin{equation}
 k^* \rightarrow 
 \left(\begin{matrix}
0 & 0 & 0 & 0 \\
0 & 1 & 0 & 1 \\
1 & 0 & 1 & 0 \\
1 & 1 & 1 & 1
\end{matrix}\right) \times
\left(\begin{matrix}
0 & 0 & 0 & 0 \\
0 & -i & 0 & 1 \\
1 & 0 & i & 0 \\
1 & 1 & 1 & -i
\end{matrix}\right) =
\left(\begin{matrix}
0 & 0 & 0 & 0 \\
1 & 1-i & 1 & 1-i \\
1 & 0 & i & 0 \\
2 & 1-i & 1+i & 1-i
\end{matrix}\right)
\end{equation} \newline The final form of eq(36) yields,
\begin{equation}\
\hat g_{\alpha \beta} =
 \begin{Bmatrix}
X_{\alpha \beta} + \Phi \partial_\alpha f \partial_\beta f & \Phi \partial_\alpha f\\
\Phi \partial_\beta f & \Phi
\end{Bmatrix}   
\end{equation} where $g_{\mu \nu}$ = $X_{\alpha \beta}$ , $\Phi$ = $X_{55}$ and hat represents a five-dimensional quantity. For any matrix X from (40)(41)(42) we could find the transpose $X^T$ and the cross product of the two
\begin{math}
\frac{1}{2}[ X \times X^T] 
\end{math}is expected to give a non-invertible matrix with determinant zero. Since $[X,X] \neq XX-XX$, the matrices are not commutative.
\subsubsection{Symplectic geometry and tautological one form}
The physically successful geometry of canonical quantization is a coordinate dependent one\cite{19}, the transformation explicitly depends on the coordinate system where the process is carried out.  The general map of a non-commutative quantization from eq(16) shows,
\begin{center}
\begin{math}
     [q^i, p_i] \xrightarrow{quantization} [q^i, -i\hbar \frac{\partial}{\partial q^i}] 
     \end{math}
     \end{center}
 \begin{center}
 \begin{math}
    \downarrow Y
    \end{math}
 \end{center}
 \begin{center}
 \begin{math}
   [Q^i, P_i] \xrightarrow{quantization} [Q^i, -i\hbar \frac{\partial}{\partial Q^i}] 
 \end{math}
  \end{center}
The symplectic form $\omega$ (exterior derivative of tautological one form $\theta$ ) is the symplectic structure,
\begin{equation}
  d\theta = \omega = dp_i \wedge dq^i  
\end{equation}
    The above equation holds canonical coordinates on manifold P and for $ M= R^5 = C^4 \times S_1 $ the wedge product yields the expansion of $\omega$ as,

    \begin{multline}
        \omega = dp_1 \otimes dq^1 + dp_2 \otimes dq^2 + dp_3 \otimes dq^3 + dp_4 \otimes dq^4 + dp_5 \otimes dq^5 - dp_1 \otimes dq^1 \\ - dp_2 \otimes dq^2  - dp_3 \otimes dq^3 - dp_4 \otimes dq^4 - dp_5 \otimes dq^5 
    \end{multline}
  The simplectic form $\omega$ is a non-degenerate form, the rest function $f$ is associated with a vector field  $X_f$ (Hamilton's vector field) such that of a category $C$ , \begin{math}
 f \in C(M) 
  \end{math} and \begin{math}
   X(f) \in X(M) 
  \end{math} through eq(17) \begin{equation}
      \omega (X_f, -) = \frac{\partial f_1}{\partial \zeta} 
  \end{equation} In coordinates, eq(46) assigns \begin{equation}
      X_{f_1} = \frac{\partial}{\partial \zeta} ( \frac{X_{\alpha \beta}}{\partial X} + \frac{X_{\beta \alpha}}{\partial X} ) -\frac{\partial}{\partial \zeta} ( \frac{Y_{\alpha \beta}}{\partial Y} + \frac{Y_{\beta \alpha}}{\partial Y} ) 
  \end{equation} This equation further yields a Poisson structure and define a property that, 
  \begin{equation}
     \prod (\partial f, \partial f_1 ) = [X_f , Y_{f}] 
  \end{equation}
  where \begin{math}
   [X,Y] = X_{\alpha \beta} Y_{\beta \alpha} -  X_{\beta \alpha} Y_{\alpha \beta} 
  \end{math} and the function $\zeta (X_f)$ represents,
  \begin{equation}
      \zeta(X_f) = \prod (\zeta) = p^i \frac{\partial}{\partial p^i}
  \end{equation} Here, in canonical coordinates \begin{equation}
 \prod = \frac{\partial}{\partial X_f} \wedge \frac{\partial}{\partial Y^f} 
  \end{equation} which we can integrate to one over all space as per assumption 1. The tautological one from of $\sigma_r^*$ and $\sigma_n^* $ commutes and integrates the space as,
  \begin{equation}
      \int  \sigma_r^* + \sigma_n^* \ [X_f , Y_f] = 1 
  \end{equation} i.e, \begin{math}
  \int L \ d(\zeta) \wedge [X_f , Y_f] \
  \end{math} gives \begin{math}
  \int \sigma_r^* + \sigma_n^* \ d(\zeta)
  \end{math} for all $\theta$ $\in$ $H$ and $f(R)$ $\in$ $H$. This is the classical action at a distance over the Lagrangian and covariant under the inverse rest function $f(R')$ of gauge symmetry,
 \begin{equation}
      \frac{\partial}{\partial \zeta} (\frac{f(R) +f(R')}{f(RR')} )
 \end{equation} Rest function for the particle pairs follows the symmetry as, \begin{equation}
  f'(R) f(R') - R'f'(RR') + Rf(RR') - \dots 
 \end{equation} where [RR’] does not commute over H which makes the function $f(R)$ deterministic and covariant in the classical limit. \section{Operationalism and observer independence}
 \subsection{On incompleteness and Einstein's argument}
 Since Spekkens showed that any $\phi$ complete models couldn't be epistemic, there arises a difficulty to show the locality of an operational theory. With the famous presupposition of wave function collapse, operationalism couldn't conceptualize measurements in structure with the independent outcomes of an experiment\cite{24},[125-157]. Further, it is also expected to associate the probability distribution with a Dirac delta function centered at the state space which is incomplete. And as a consequence, if the operational theory partially determines ‘what is being observed’, observers holding incompatible theories means that the different epistemic outcomes which are irrelevant in the narrow sense of structurally compatible models are presuppositions as argued in the Bell's theorem. Supporting compatible model held by Einstein prompts that the quantum state aligned to a system might goes away in any direction (representing what reality it offers and depending on the system of interest) when interacted with the part of any non-local environment. Interestingly, there supposed to be two separate states of preparation where Einstein’s argument is that “quantum mechanical description of physical reality is not complete\cite{25} but epistemic” by the virtue of hidden variable theory, a considerable case that can be discussed.
\begin{enumerate}
\item There is no prominent need of Bell’s argument to show the non-locality of experimental results coming out of an operational theory, if $\phi$ is ontic. Bell’s argument is prominent only if we attempt ruling out any locality assumptions of hidden variable theories.
\item If $\phi$ is not ontic, then it is called $\phi$-incomplete. Here, there also require a framework on how such a model is incomplete and require a strong philosophy to interpret the theory in more of an operational manner.
\item An ontic state space $\Lambda$ is precisely the projective Hilbert’s space where $\Lambda = PH$.
\end{enumerate}
It is valid that in this sense, there also arise another difficulty
to understand the measurement problem from the unique experimental outcomes of supplementary assumptions\cite{22}. Significantly to a greater extent, unlike any classical theory, the  analysis is predicated on the assumption that, of the potential outcomes of a given experiment, one and only one outcome occurs, and hence it makes sense to speak of the outcome of an experiment.
\subsection{Quantization of rest function and hidden value assumption}
Suppose if there are two non-interacting particles where none of them is observed at first and $\nabla$ is the partial measurement operator, then for the wave function $\phi$ , the spin eigen functions are in the form,
 \begin{equation}
     \langle \phi_1 \phi_2 \rangle  +   \langle \phi_2 \phi_1 \rangle = \sqrt{\frac{1}{2}} \nabla \phi 
 \end{equation}
 For Hamilton $H_e$ the Kronecker product $*$ gives, \begin{equation}
     \mid \phi_1 \rangle \ * \ \mid \phi_2 \rangle \ + \mid \phi_2 \rangle \ * \ \mid \phi_1 \rangle = \sqrt{\frac{1}{2}} \langle H_e \frac{\partial \phi}{\partial t} \rangle 
 \end{equation} Immediately after the measurement $\nabla$ is assumed to be a covariant form of function $\phi$ such that the equation, \begin{equation}
      \nabla_\phi = i \hbar \frac{\partial \phi}{\partial t} + H_e \phi 
 \end{equation} is Schrodinger's time dependent wave equation. For the function $\phi$ we could recast in polar form as,
 \begin{equation}
    \phi (r,t) = \sqrt{R(r,t)} \ exp \frac{iS(r,t)}{\hbar} 
 \end{equation} Where,
 \begin{center}
 \begin{math}
      r = \sqrt{2 \pi \sqrt{\frac{r}{iS}}\ log_{\zeta_i}(\frac{\partial f_{1}}{\partial_{\zeta}})} 
 \end{math}
 \end{center} The simplectic form of $\omega$ could be expressed in terms of r from eq(46) and intuitively yields an invariance  of rest function $f(R)$ deterministic under the transformation of elementary nods $\zeta_i$ which is a logarithmic function. Since $\zeta_i$ is a time evolving radical function in Schrodinger equation, there also confronts the measurement paradox of quantum observables $\nabla$ that for every bonds of $\zeta_i$ there exists a reduced variable symmetry with the respect to the evolution of wave equation independent of time.
 Following Bohm, we could substitute eq(57) into generalized Klein-Gordon eq(6),
 \begin{equation}
  [\delta_{i} \delta^{i}- m^2-\zeta (R)]\sqrt{R(r,t)} \ exp \frac{iS(r,t)}{\hbar} = 0  
 \end{equation} Here, localizing eq(58) could gives us two possible solutions, where the second term of LHS could never be zero. The only possible term to be zero gives us the solution of elementary position $\zeta (R)$ \begin{equation}
     \delta_{i} \delta^{i}- m^2 = \zeta (R) 
 \end{equation}
 \begin{equation}
     \zeta (R) = \frac{\hbar^2}{m^2} \frac{\nabla R}{R} 
 \end{equation} Intuitively $\zeta (R)$ is the quantum potential $Q$ of a Bohemian system. But eq(59) is in conflict with relativity since the rest mass $m$ depends upon the quantum potential $Q$ and not a constant in reference frame $s(x)$. For such reasons, in a five-dimensional manifold eq(60) can be modified as,
 \begin{equation}
     \zeta (R) = \frac{iS}{m^2 r^3} \frac{\nabla R}{R} 
 \end{equation} And substituting for $r$ we get,
 \begin{equation}
       \zeta (R) = \frac{iS(r,t)}{m^2 (2 \pi \sqrt{\frac{r}{iS(r,t)}} log_{\zeta_i} \omega (X_f, -))^{\frac{3}{2}}}  \frac{\nabla R(r,t)}{R(r,t)} 
 \end{equation}
 In eq(62) the logarithmic function of elementary nodes depends upon the elementary position $\zeta (R)$ where the mass m depends upon the rest position of any one of the particle pairs. In such a way, the principles of relativity are never violated in a four-dimensional manifold and behave quantum mechanically in a five-dimensional scalar  field. \newline
 Now, for the wave equation $\phi (r,t)$ there are two discrete possibilities, \begin{enumerate}
     \item $\phi (r,t)$ as an observable function of r under the operator $\nabla_\phi$ or
     \item   $\phi (r,t)$ as an independent function of clock time $t$.
 \end{enumerate}
 The second possibility may contradict the energy-momentum conservation from assumption 1 that for all discrete values, the spin of the particle exists as  -1/2 or  1/2 , 2/3 or 1/3 , etc. as real values. For all definite entangled pairs under the partial measurement operator $\nabla$, there only exists the possibility 1 that for any observable function independent of nodes $\zeta_i$ the spin eigenvalues would also turn independent of the real value representation of $R$ and $S$ over the topology of Hilbert's space $H$.\newline The incompatibility of observables to regard spin under the real vale representation of $\phi (r,t)$ depends on elementary position $\zeta (R)$ with a hidden value assumption, such that spins of real values (1/2 , -1/2 ) never contradicts assumption 1 before measurement.
 let, 
 \begin{equation}
     \langle \phi_1 \mid \phi^\dagger \rangle \ \langle \phi_2 \mid  \phi^\dagger \rangle = \langle \phi_1 \mid  \phi \mid  \phi_2 \rangle
 \end{equation} Here, $\dagger$ represents the conjugate function. Such a wave function should take the prejudice of time-symmetry before measurement. But this form cannot be localised with a real value coming out of a hidden value assumption. So we implement another function $\psi$ irrespective of the source function $\phi$. \newline
 Here, $\psi$ evolves under Schrodinger's time independent wave equation, \begin{equation}
      i\hbar \frac{\partial \psi}{\partial t} = \psi [-\frac{\hbar^2}{2m} \nabla^2 +R] 
 \end{equation} Which can be rewritten as,
 \begin{equation}
       i\hbar \frac{\partial \psi}{\partial t} = \psi \ (K.E + \oint r \delta x \wedge [X_f , Y_f] )
 \end{equation} where first term of RHS is the kinetic energy and second term is the Hamilton-Jacobi action extended over the path. The partial measurement regardless of spin over the entangled pairs gives the form,
 \begin{equation}
     \frac{\langle \psi_1 \mid  \psi \mid  \psi_2 \rangle \langle \psi_1 \mid  \psi\mid  \psi_2 \rangle^\dagger }{\parallel \sqrt{ \langle \psi_1 \mid  \psi \mid \psi_2 \rangle }\parallel} 
 \end{equation} where factorizing $\langle \phi \mid \psi \rangle $ gives,
 \begin{equation}
   \parallel \frac{\langle \phi \mid \phi \psi \mid \psi \rangle }{\sqrt{\phi \psi}} \parallel 
 \end{equation} Since $\phi$ being orthogonal to trace elements, we get only
 \begin{equation}
         \frac{\langle \phi_1 \mid  \phi \mid  \phi_2 \rangle \langle \psi_1 \mid  \psi\mid  \psi_2 \rangle^\dagger }{\parallel \sqrt{ \langle \phi_1 \mid  \phi \mid \phi_2 \rangle}\parallel} \geq 0
 \end{equation} And eq(67) turns out to be an independent function of time. So there only exists a minimal possibility for function $\mid \phi \rangle$ to commute with $\mid \psi \rangle$ before measurement. This feature turns out to be what we call as entanglement where neither of the wave function could be measured at first. If the hidden value assumption is valid, then
 \begin{equation}
      P(f([\sigma_\phi , \sigma_\psi])) \leq 1 
 \end{equation} For minimal possibilities, such that for spin, the probability of $\psi_1$ to have a spin $\uparrow$ is 1/2 and $\psi_2$ to have a spin $\downarrow$ is -1/2. If that is the case, then
 \begin{equation}
P(\langle \pm \frac{1}{2} \rangle ) \geq f(\langle \phi \mid \psi \rangle) 
 \end{equation} Here $\langle \phi \mid \psi \rangle$ is the real value after measurement. From eq(69) and eq(70), it further turns out to be,
 \begin{equation}
   f([\sigma_\phi , \sigma_\psi]) \leq P(\langle \pm \frac{1}{2} \rangle ) 
 \end{equation} The above equations shows that \begin{math}
  f[\sigma_\phi, \sigma_\psi] 
 \end{math} have the maximum possibility for the spin to be $\uparrow$ or $\downarrow$ where the option experienced by the observer to have a "yes" or "no" exists nevertheless there is a minimal possibility of \begin{equation}
      P( f([\sigma_\phi , \sigma_\psi]) ) \geq P(\langle \uparrow +\frac{1}{2} , \downarrow -\frac{1}{2} \rangle) 
 \end{equation} for all non-abelien group that belongs to the lie group of $\phi(r,t)$. The violation of bond set by eq(72) gradually tends to violate the hidden value assumption since \begin{equation}
f(\zeta (R)) \geq P(\langle \frac{1}{2} \rangle) 
 \end{equation} where $f(\zeta (R) )$ is a function of finite possibilities. Even after measurement $\mid \psi \rangle$ holds the minimal possibility for function $\langle \psi_1 \mid \psi \mid \psi_2 \rangle$ normalized with state vectors of spin $\uparrow$ and $\downarrow$ with consecutive hidden values. The assumption is true for all entangled pairs if $\mid \phi_1 \rangle$ commutes for real values of $\frac{1}{2} \langle \psi_1 \mid  \psi \mid  \psi_2 \rangle$ and for hidden values of $\frac{1}{2} \langle \psi_2 \mid  \psi \mid  \psi_1 \rangle$. Since $\phi (r,t)$ holds the eq(68) and eq(73), the hidden value assumption is not valid for all $\mid \phi \rangle \mid \psi \rangle$ $\in$ $PH$. 
 \subsection{Categories and the structure of operational theory}
From classical assumptions, a category C is defined as $(\lambda_1 \oplus \lambda_2) \otimes Agent \simeq (\lambda_1 \otimes Agent ) \oplus (\lambda_2 \otimes Agent) $ where $\otimes$-structure capture the notion of entanglement and $\oplus$-structure capture the notion of matrix calculus. But these two are not independent and persists a discrete outcome independent from the structure of $R$. Let $R \subseteq R_1 \times R_2$ consisting of entanglement pairs satisfying the relation,
\begin{equation}
    R \mid (\boldsymbol{rel} , \times ) : R \times R :: \lambda_1 \xleftarrow{\pi_1} \lambda_1 \lambda_2 \xrightarrow{\pi_2} \lambda_2
\end{equation} where R is a group of all topological spaces $R_1 \otimes R_2$ and $\lambda_1 ,\lambda_2$ is the associated epistemic values for all $R_1 , R_2 \in C$. If $R$ act like a locally small group of all possible outcomes, then for a $\psi$ epistemic model, the characteristic function $\pi_1 : 1 \mapsto (Yes, No)$ of $(R \mid \lambda_1 \leq \boldsymbol{rel} \ R_1 )$ and $\pi_2 : 0 \mapsto (Yes, No)$ of $(R \mid \lambda_2 \leq \boldsymbol{rel} \ R_2)$ yields a monotone map if (0,1) is given a standard ordering of pairs $(0 < 1 )$. Hence, R is not surjective.
\begin{center}
   $ C \mid \sum_\mu \mid : X_{\alpha} R(+, -)$
\end{center}
\begin{center}
 $\forall_{\alpha} :-  \ \ X \longrightarrow(H_1 , H_2 , \dots H_n )$
\end{center} for a representable functor $X_\alpha$. The independency of outcomes can be expressed in operational terms by characterising $\sum_\mu$ as the trace operator 
\begin{equation}
\sum_\mu H_\mu = I
\end{equation} and contains the co-linear functions of $R_1$ and $R_2$ as
\begin{equation}
\begin{split}
&\forall_{f,g} \\
& \ \ \ \ \ \ \ \mid R_1 \rangle \langle R_2 \mid \oplus \mid R \mid =  2 \pi  \sum_\mu H_\mu
\end{split}
\end{equation}  Since picking more than two elements from $R$ is restricted by the presuppositions of operational theory, two observers Alice and Bob only yields,
\begin{center}
$Yes:  \langle + \mid - \rangle \longrightarrow (P(Yes \mid 0, 1 \rangle) \oplus P (No \mid 1, 0\rangle)) \otimes P(Yes,No \mid \lambda_1) d\lambda_1 = P(\langle \pm \frac{1}{2} \rangle )$ 
\end{center} and
\begin{center}
$No:  \langle + \mid - \rangle \longrightarrow (P(No \mid 1, 0 \rangle)  \oplus P (Yes \mid 0, 1\rangle)) \otimes P(Yes, No \mid \lambda_2) d\lambda_2 = P(\langle \pm \frac{1}{2} \rangle )$ 
\end{center} 
by swapping all possible $\boldsymbol{rel}$ :- $\langle 00 \rangle , \langle 01 \rangle, \langle 10 \rangle, \langle 11 \rangle$. 
Bob's element of R shall be independent of Alice's element of R but also induces a unique function over the diagonal of $\boldsymbol{\Lambda}$ as $R \longrightarrow R_1 \otimes R_2$ that arises in  the light of functors $f$ and $g$.
\begin{equation}
P( Yes \mid yes, no ) := f: R \longrightarrow  \boldsymbol{\Lambda} \xleftarrow{Alice} R_1 \otimes R_2 \xrightarrow{Bob} \boldsymbol{\Lambda}
\end{equation}
 and
\begin{equation}
P( No \mid yes, no ) := g: R \longrightarrow  \boldsymbol{\Lambda} \xleftarrow{Bob} R_1 \otimes R_2 \xrightarrow{Alice} \boldsymbol{\Lambda}
\end{equation} From the outcome Independence of eq(37) and eq(39) we could substantiate the result yield by Alice as,
\begin{equation}
\boldsymbol{\Lambda}:-   P(Yes \mid R \rangle )  \simeq P(Yes, No) d\boldsymbol{\Lambda} \geq P\langle \frac{1}{2}\rangle
\end{equation} and 
\begin{center}
  $X: \boldsymbol{\Lambda} \longrightarrow R_1 \otimes R_2$
\end{center} where eq(41) and eq(42) is the $\langle j^{*}_+, k^{*}_ - \rangle$ ontic correspondence of any operational theory for Alice which necessarily allows her to represent $X$ if there any complete category C. This is no different for Bob, when he could pick an element from $R$ and compare his results with that of Alice.  

\section{BPF tensor field algebra}
To avoid the complexity of equations arriving at the BPF scalars, there are a set of assumptions that I have to make. From eq(35)(53) and (62) it is know that, \begin{enumerate}
     \item $\zeta (R)$ is a real value function of complex dynamical system and invariant topological operations.
     \item $\phi(t)$ covariantly represents the clock-time interval of the wave $\nabla(\tau, \phi_1, \phi_2, \\ \dots \phi_{n-1} )$.
     \item $f(R,t)$ is the void-generating function given that the wave function $\phi$ does not allow any super-luminial communication at $n \leq 4$.
     \item Voids at BPF represents the vacuum energy which is an invariant functor.
 \end{enumerate}
 With these conditions, eq(35) turns into
 \begin{equation}
 \begin{split}
         f_1(RR')_k \nabla \phi + \frac{i^2}{2} f(\zeta(R)) - \zeta_{jk} \int f(\zeta)_i \ g_{jk} =  \frac{2m}{C^2} \int  \frac{G}{2} \sqrt{\frac{iS}{r}}+ \ [\epsilon_r \epsilon_n] \ F [\sigma_r ,\sigma_n]
 \end{split}
 \end{equation} 
 To solve the above equation we need to split the LHS in to three parts and solemnly resolve individual parts. Empirically verifying each parts of the LHS may have different physical significance and to reduce the complexity we only consider such.\vspace{5mm} \newline 
 Part 1 :  For dummy indices \ $\alpha \beta \gamma$, extending $R$ and $f$ for $n-(m-1)$ dimensions we get,
 \begin{equation}
 \begin{split}
         R_{\alpha \beta}^\gamma R_{\gamma}^{\beta \alpha} - f_1 R_1 (R_{\beta \alpha}^\gamma R_{\alpha \beta}^\gamma ) + f_2 R_2 (R_{\alpha \beta}^\gamma R_{\beta \alpha}^\gamma ) \\ - f_3 R_3 (R_{\beta \alpha}^\gamma R_{\alpha \beta}^\gamma ) + \dots \ - f_{n} R_{m-1} (R_{\beta \alpha}^\gamma R_{\alpha \beta}^\gamma )  
 \end{split}
 \end{equation} For five dimensions the transformation yields the result,
 \begin{center}
 \begin{math}
       \rightarrow R_{\alpha \beta}^\gamma  \ \wedge \ -R_{\beta}^{\gamma \alpha} 
 \end{math}
 \end{center} Multiplying the result with $R$ takes the form,
 \begin{center}
 \begin{math}
       R^\gamma \otimes R_{\alpha \beta} \ \cdot \ R^{\alpha \gamma} \otimes  R_{\beta} \ \rightarrow \ R_{\alpha} R_{\beta}^\gamma \ \cdot \ R^{\alpha} R_{\beta}^\gamma 
 \end{math}
 \end{center} Rewriting the above equation in the prescribed form $f(RR')$ there only exists the scalar \begin{equation}
      \zeta R 
 \end{equation}
 {\bf Remark}: Action at the Planck’s effective $r$ is not a scalar product of energy-momentum tensor equation typically does for the classical field equations. \vspace{5mm} \newline
 Part 2: For breaking the gauge symmetry using generating function f we have gravitational metric tensor g,
 \begin{center}
 \begin{math}
      \frac{i^2}{2} f(RR') \ g(RR')  
 \end{math}
 \end{center} which yields
 \begin{equation}
 \begin{split}
       \rightarrow  \frac{i^2}{2} [f(g(\phi)) -g(f(\phi))] \ + \ i^2 [f'(g'(\phi))-g'(f'(\phi))] \\ - \frac{3i^2}{2} [f''(g''(\phi))-g''(f''(\phi))] \ + \dots 
 \end{split}
 \end{equation} For all $[f,g]_\phi$ ; $\phi(t)$ dominate over the time series with an increasing space exponential term,
 \begin{center}
 \begin{math}
            R\ exp({\frac{i^2}{2}}) ( \ [f, g] - R'[f', g'] + R'' [f'', g'']- \dots ) 
 \end{math}
 \end{center} that gives, 
 \begin{equation}
    R\ exp(\frac{i^2 \hbar}{4 \pi}) 
 \end{equation} 
 as the quantum length generator obtained from the power series of $\zeta (R)$ elementary position commuted over a distance r.\vspace{5mm} \newline
 {\bf Remark}: At the sub-quantum level, the metric $g$ is defined as a generating  function of rest potential $f(\zeta(R))$ and not the Planck's effective curvature, which makes it non-commutative and synthetic over the BPF scalars.\vspace{5mm}
 \newline
 Part 3: Under a radical time symmetry, the BPF tensor variables commutes as,
 \begin{equation}
     \zeta_{jk} \int f(\zeta)_i \ g[R_{ij}^k \ R_{jk}^i] 
 \end{equation} which yields the form,
 \begin{center}
 \begin{math}
    \zeta_{jk} \int \frac{\zeta_i f_1}{\zeta^2} + \frac{\zeta_i f_2}{\zeta^2} + \frac{\zeta_i f_3}{\zeta^2} + \dots  g_m [R_{ij}^k \ R_{jk}^i] 
 \end{math}
 \end{center} Further gives
 \begin{center}
 \begin{math}
       \zeta_{jk} \int \frac{\zeta_i f_1}{\zeta^2} \ g_1 [R_{ij}^k \ R_{jk}^i] \ + \zeta_{jk} \int \ \frac{\zeta_i f_2}{\zeta^2} \ g_2 [R_{ij}^k \ R_{jk}^i] +  \zeta_{jk} \int \frac{\zeta_i f_3}{\zeta^2} \ g_3 [R_{ij}^k \ R_{jk}^i] \ + \dots \zeta_{jk} \int \frac{\zeta_i f_{n-1}}{\zeta^2} \  g_m [R_{ij}^k \ R_{jk}^i] 
 \end{math}
 \end{center}
 Rewriting the above equation in the form \begin{math}
 \frac{ \zeta_{jk}}{\zeta^2} \ [f, g] 
 \end{math} we get, \begin{equation}
  \frac{\zeta_{jk}}{\zeta} 
 \end{equation}
 Now, combining the three parts i.e, eq(82) (84) and (86) we get a five dimensional field as \begin{equation}
      \zeta R + R\ exp(\frac{i^2 \hbar}{4 \pi}) - \frac{\zeta_{jk}}{\zeta}  =  \frac{2m}{C^2} \int  \frac{G}{2} \sqrt{\frac{iS}{r}}+ \ [\epsilon_r ,\epsilon_n] \ F [\sigma_r ,\sigma_n] 
   \end{equation} And comparing with eq(7) we get 
   \begin{equation}
         \zeta \hat R + \hat R\ exp(\frac{i^2 \hbar}{4 \pi}) - \frac{\zeta_{jk}}{\zeta} = - \zeta \ \hat T^{jk} 
   \end{equation} where hat represents a five dimensional quantity and $\zeta = 2 \pi \mu G$ is the scalar constant of BPF invariant scalar field.
\section{Pregeometric significance of BPF scalar }
Pregeometry traditionally described by design is to account for a space-time geometry resting on entities ontologically prior to it and of an essentially new character\cite{20}. However, it turned out that the distinctive feature of pregeometric approaches confronts the assumption of space-time being precisely manifested by a pseudo-Riemannian manifold; only to replace it with some type of geometric-theoretical assumption. \newline 
As per J.A. Wheeler, the inventor of this approach, there are two important pronouncements \begin{enumerate}
\item A concept of pregeometry that breaks loose at the start from all mention of geometry and distances.
\item The approach of distance to give up on the search for pregeometry structure of space and time.
\end{enumerate}
I can exemplify from the preceding statements that Wheeler was looking for a non-geometric structure of space and time. In fact, the need of a fundamental theory independent from topology and dimensionality is at the heart of Wheeler's statement. Deeply, I can also regard his statements as a fundamental proposition of quantum theory that gives rise to the idea of pregeometry; pregeometry makes geometry; geometry give rise to matter, physical laws
and constants of the universe. \vspace{4mm} \newline
There are few outcomes that I have come up with in order for a wave function $\phi$ to posses a particular feature like entanglement in the pregeometric structure of elementary nodes. The first among them is the Planck's effective field free from the influence of mass. According to the diffeomorphism invariance principle of General relativity, there exists no physical quantity as a possible feature of empty space but only available as a quality coming out from the influence of fields. In other words, no space is compatible in the absence of fields. Because of this pattern, there are physical incompatibilities coming out of a scalar field apparently with no physical mass, from eq(88) and eq(32) we have,
\begin{equation}
\zeta \hat R + \hat R\ exp(\frac{i^2 \hbar}{4 \pi}) - \frac{\zeta_{jk}}{\zeta} = \ Fid_\epsilon [\sigma_r ,\sigma_n] \geq \frac{\hbar}{2}
\end{equation} where R could take any real value. This proposition further yields that the effective Planck's wavelength $r$ of a photon is not certain at $n=5$ dimensions since the exponential curve is light like with any increase or decrease in the value, the fundamental assumption of $\mu \neq 1$ or $R=0$ would give a transition like
\begin{equation}
\zeta_{jk} \geq \ \pi \mu G \hbar
\end{equation} 
which gives a value of \ $8.43070669 \times 10^{-99} \ J m^3 s^{-1}$ as the minimal energy required for two nodes structuring like a cube. As expected by Requadart and Roy, if $\zeta_{jk} = \pm 1$ or 0 one can draw a directed bond, $b_{ik}$, if $\zeta_{jk} = +1$ , with $\zeta_{jk}$ = $-\zeta_{kj}$ implied and delete the bond if $\zeta_{jk}$ = 0. Moreover the geometric structuring is proportional to $\mu = \frac{m_e^2}{m_p}$ where no bonds could be formed in a light-like curve of $\mu$ = 0. Or in other words, for a photon with rest mass zero, fundamentally $\mu \neq 1$ and there is no possibility of a curvature with positive value of R. Even though R exists in a physically perceivable time-space quanta as an observable function of $\phi$ at $r,n \leq 4$ by any means, the entangled structure of nodes are not materialized to any physical tests or embracements by the diffeomorphism invariance of BPF scalar. \vspace{4mm}\newline
Manfred Requardt in his paper\cite{21} have argued that the apparent non-locality of physical universe is coming from the two-storey structure of the medium, what he call as a web of lumps. These lumps form a structure with surface ST where the initial phase is described as the physical vacuum that expresses itself in the complex superposition of momentary waves of ordinary quantum model QX. The particular phase of medium QX/ST as considered by Requardt is a language of dynamical systems in a wider context of complex systems. In the BPF scalar model, assumption 2 and assumption 3 are in the similar context of the locality and better understood by the prescribed level of overlapping and non-overlapping lumps. The effective Planck's wavelength from assumption 2 triggers the path of inter-nodes irrespective of the distance between any two particles where information is commuted instantaneously via these networks. As expected by Requardt, the information passed between the rest position of two points over BPF form the remnants of the scale $n$ for the existence of the quantum/scalar potential of assumption 3. The poly dimensional approach of BPF scalar shows that the feature (entanglement) of particle pairs could be well understood from the local behaviour of four dimensional manifolds of ordinary space-time. The discrete nature of manifolds could also be verified using normalized nodes in projective Hilbert's space. The respective approach to show the instantaneous communication between two discrete spaces are not emerging out of dimensionality, rather as a consequence of losing one. This can be clarified by a dimensional analysis of $\zeta_{jk}$. It is also under the domain of non-locality that, if the notion of distances and time gets vanished away at a coarse grained level, out of non-dimensionality, infinite regions of ordinary space-time may act like a finite region in Hilbert's space.

\section{Conclusion}
A prominent discovery from Bell's experiment showed that two distant parts of an entangled system know more about each other than the local. Rethinking space-time, a toy model of Planck's scale physics is proposed. From three underlying assumptions that are used to construct the model, I found the use of natural geometry produces an insignificant description beyond Planck's scale. The direct consequence of pseudo-Riemannian curvature achieved from the theory obtain an abstract representation of how the dynamics of interconnected nodes contribute to the feature of entanglement. This can be significantly explained with Requardt-Roy cellular network model with sufficient consideration of Beyond Planck's Field. Such a description suggests the geometric effects at four dimensions may attribute pregeometric effects at five dimensions where the best consequence being gravity.

 \end{document}